\documentclass[12pt]{article}

\usepackage{adjustbox}
\usepackage{amsfonts,amsmath,amssymb, amsthm}
\allowdisplaybreaks
\usepackage{appendix}
\usepackage{bm}
\usepackage{braket}
\usepackage{xcolor}
    \definecolor{darkgreen}{rgb}{0,0.5,0}
    \definecolor{darkred}{rgb}{0.5,0,0}
    \definecolor{darkblue}{rgb}{0,0,0.6}
    \definecolor{purple}{rgb}{0.4,.2,0.7}
\usepackage{comment}
\usepackage{esint}
\usepackage{euscript}
\usepackage{float}
\usepackage{graphicx}
\usepackage[hyperfootnotes = true, colorlinks = true, linkcolor = darkblue, citecolor = purple]{hyperref}
\usepackage{import}
\usepackage{makecell}
\usepackage{multirow}
\usepackage{subcaption}
\usepackage{tcolorbox}
\usepackage{tikz}
\usepackage[normalem]{ulem}

\def\ZZZ{{\hbox{ Z\kern-1.6mm Z}}}
\def\RRR{{\hbox{ R\kern-2.4mm R}}}
\def\CCC{{\hbox{ C\kern-2.0mm C}}}
\def\zzz{{\hbox{z\kern-1mm z}}}
\def\ZZZ{\mathbb{Z}}
\def\RRR{\mathbb{R}}

\newcommand{\qeq}{{\hbox{=\kern-2.3mm ? \kern.5mm }}}
\renewcommand{\qeq}{=}

\newcommand{\eps}{\epsilon}

\newcommand{\GG}{{\cal G}}

\newcommand{\HH}{{\cal H}}
\newcommand{\MM}{{\cal M}}

\newcommand{\OO}{{\cal O}}

\newcommand{\LL}{{\cal L}}

\newcommand{\wt}{\widetilde}

\newcommand{\NN}{{\cal N}}

\newcommand{\SSS}{{\cal S}}

\newcommand{\be}{\begin{equation}}
\newcommand{\ee}{\end{equation}}
\newcommand{\ben}{\begin{eqnarray}\displaystyle}
\newcommand{\een}{\end{eqnarray}}
\newcommand{\refb}[1]{(\ref{#1})}
\newcommand{\p}{\partial}
\newcommand{\sectiono}[1]{\section{#1}\setcounter{equation}{0}}

\def\one{{\hbox{ 1\kern-.8mm l}}}
\def\zero{{\hbox{ 0\kern-1.5mm 0}}}

\newcommand{\bea}[1]{\begin{eqnarray}\label{#1} }
\newcommand{\eea}{\end{eqnarray}}

\numberwithin{equation}{section}

\usepackage[margin = 2.2cm]{geometry}
\usepackage[ragged]{footmisc}
\begin{document}

\thispagestyle{empty}
\begin{center}
    ~\vspace{5mm}
    
    {\Large \bf 

		Temperature and Moduli Independence of the Supersymmetric Index from Gravity
        
    }
    
    \vspace{0.4in}
    
    {\bf Raghu Mahajan and Ashoke Sen}

    \vspace{0.4in}

    International Centre for Theoretical Sciences, Bengaluru, Karnataka 560089, India
    
    \vspace{0.1in}
    
    {\tt raghu.mahajan@icts.res.in, ashoke.sen@icts.res.in}
\end{center}

\vspace{0.4in}

\begin{abstract}

% Original abstract (commented out 2026-09-24):
% The index of supersymmetric black holes can be computed directly from the gravitational path integral. 
% The index is expected to be independent of the temperature and the asymptotic values of the scalar moduli. 
% This has been verified by explicit computations in many cases,   
% but a general proof is lacking even at the classical level. 
% [Version before the 2026-09-24 edit:] We fill this gap by showing that in any four-dimensional flat-space supergravity theory with two or more supersymmetries,  the classical contribution to the index is independent of the temperature and the asymptotic values of the moduli.
% We fill this gap by showing that in any four-dimensional supergravity theory with $\mathcal{N}\geq 2$ supersymmetry in asymptotically flat space-time, the classical contribution to the index from each supersymmetric saddle is independent of the temperature and of the asymptotic values of the moduli. 
% Our results hold in the presence of higher derivative terms and for multi-centered black hole configurations.

The supersymmetric index, which gives a signed count of the microstates of black holes, can be computed directly from a gravitational path integral with appropriate boundary conditions.
From general arguments based on the supersymmetry algebra,  the index is expected to be independent of the temperature and,  away from walls of marginal stability,  of the asymptotic values of the scalar moduli.
On the gravity side, this has been verified by explicit computations in many examples, but a general proof is lacking even at the classical level.
We fill this gap by showing that in any four-dimensional supergravity theory with $\mathcal{N}\geq 2$ supersymmetry in asymptotically flat space-time, the classical contribution to the index from each supersymmetric saddle is independent of both.
The argument uses only the behavior of the solution and of the Killing spinor equations near infinity.
The result therefore holds in the presence of arbitrary higher-derivative corrections and for multi-centered black hole configurations.
In particular, moduli independence follows without appeal to the attractor mechanism, which relies on the near-horizon geometry.

\end{abstract}

\pagebreak

\tableofcontents

\sectiono{Introduction}

One of the major successes of string theory has been the demonstration that, for
certain classes of supersymmetric black holes, one can match the microscopic entropy
\cite{Susskind:1993ws,Sen:1995in,9601029} with the macroscopic entropy
\cite{Bekenstein:1973ur,Hawking:1975vcx,Gibbons:1976ue}. 
One of the weak points in the analysis was
that the microscopic analysis was done at weak coupling and the macroscopic analysis
was done at strong coupling. This could be avoided by computing an appropriate
index \cite{Witten:1982df,Ferrara:1981qd,Bachas:1996bp}, which is protected under a change in the coupling constant, instead of the degeneracy.
See \cite{Mandal-Sen} for a review.
However,
while on the microscopic side one could compute the index, the 
Bekenstein--Hawking--Wald entropy 
\cite{Bekenstein:1973ur,Wald:1993nt} on the macroscopic side describes the logarithm
of the degeneracy and not that of the index.

This problem was finally resolved by finding a way to compute the index directly on the
macroscopic side by a gravitational path 
integral \cite{1810.11442,2107.09062}. 
This prescription modifies the Gibbons--Hawking path integral \cite{Gibbons:1976ue}
by including a chemical potential conjugate to angular momentum $J_3$ and choosing its value
so that the trace over states includes the insertion $\exp(2\pi i J_3)$.
The insertion equals $+1$ on bosonic states and $-1$ on fermionic states.
Writing $F$ for fermion number, it therefore equals $(-1)^F$ and converts the degeneracy to an index. Many tests of this
formula have been performed, both at the 
classical \cite{1810.11442,2107.09062,2310.07763,2411.08260,2511.18771} 
and at the 
quantum \cite{2306.07322,2605.30417,2606.17149}
level. In this paper, we shall restrict our analysis to the classical level.

We now give a quick summary of the prescription for computing the classical
index before we state the problem that we address. 
We shall restrict our analysis to four space-time dimensions and to the asymptotically flat case, although we expect
the results to be valid in other dimensions and in the asymptotically AdS case as well.
We compute the path integral
over fields in Euclidean space with the following boundary conditions. 
Let $\tau$ be the Euclidean time coordinate. We make the identification
\be\label{e1}
(\tau,\phi) \equiv (\tau+\beta, \phi+i\omega)\, .
\ee
Eventually, we shall set $\omega=-2\pi i$.
We also impose the following asymptotic
boundary condition on the metric $g_{\mu\nu}$, the gauge field $A_\mu^{(i)}$
and the scalar moduli fields $\phi^{(a)}$:
\be \label{e2}
g_{\mu\nu}=\delta_{\mu\nu}, 
\qquad A_\tau^{(i)}=-i \, \nu^{(i)}/\beta, \qquad \phi^{(a)} = m^{(a)}\, .
\ee
We also fix the boundary condition on the angular components of $A^{(i)}_\mu$ by
requiring that 
\be\label{e1.3}
\int F^{(i)} = P^{(i)}\, ,
\ee
where the integral is carried out over the celestial sphere and $F^{(i)}=dA^{(i)}$.
Thus,  the $P^{(i)}$'s are equal to the magnetic charges.

For $\omega = -2\pi i$, the 
path integral receives contributions from some supersymmetric classical solutions.
These can be described as the Euclidean continuation of a rotating dyonic black hole
solution with imaginary angular momentum,  or its multi-center generalizations \cite{2310.07763,  2507.07166}.
Let $I$ be the Euclidean action for a solution with the asymptotic boundary conditions
given above. Following \cite{Gibbons:1976ue,9607108}, we define
\be \label{e3}
M := {\p I\over \p\beta}\bigg|_{\beta, \nu,\omega,m}, \qquad Q_{(i)} := 
{\p I\over \p\nu^{(i)}}\bigg|_{\beta, \nu,\omega,m}, \qquad 
J := {\p I\over \p \omega}\bigg|_{\beta, \nu,\omega,m}, \qquad 
\Sigma_{(a)} := {1\over \beta} {\p I\over \p m^{(a)}}\bigg|_{\beta, \nu,\omega,m}\, ,
\ee
where the symbol $\bigg|_{\beta, \nu,\omega,m}$ means that, while taking the derivative,
we keep fixed every variable in the list other than the one with respect to which we are
differentiating. 
The magnetic charges $P^{(i)}$ are always kept fixed, and so we do not write them explicitly. 
We also define
\be \label{e4}
S_{\mathrm{BPS}} := - I + \beta M + \nu^{(i)}\, Q_{(i)}\, .
\ee
The arguments of \cite{1810.11442,2107.09062} 
show that $S_{\mathrm{BPS}}$ computes the logarithm of the desired
index at the leading order. 

Now, to begin with, $I$ is a function of the variables $P^{(i)}$, $\nu^{(i)}$, $\beta$ and
$m^{(a)}$. 
We could eliminate $\nu^{(i)}$ in favor of $Q_{(i)}$ using the second equation in \refb{e3}. 
In that case, $I$, and hence $S_{\mathrm{BPS}}$, would be functions of  
$P^{(i)}$, $Q_{(i)}$, $\beta$ and $m^{(a)}$. 
With this choice of variables,  microscopic results for $S_{\mathrm{BPS}}$ are manifestly independent of $\beta$ and $m^{(a)}$,  as follows from general arguments based on supersymmetry.
On the macroscopic or gravity side,  all explicit computations show that $S_{\mathrm{BPS}}$ is indeed independent of $\beta$ and $m^{(a)}$,  but there is no general proof.\footnote{Earlier arguments for moduli independence of the index relied on the attractor mechanism \cite{9508072,9602111,9602136,9603090}, which uses the near-horizon geometry.}

The goal of this paper is to give a general proof of the $\beta$ and $m^{(a)}$ independence of $S_{\mathrm{BPS}}$ in any theory with $\NN\ge 2$ supersymmetry in four space-time dimensions,  without making any assumptions on the higher-derivative
terms that the theory might have. 
Our proof also applies to multi-center solutions.
However, we do need to assume the existence of
a classical solution satisfying the boundary conditions \refb{e1}, \refb{e2}, \refb{e1.3}
and preserving at least one supersymmetry; without this, the statement of
the problem itself becomes meaningless. 
We summarize the general strategy
of the proof in section \ref{estrategy} and fill in the details in the rest of the sections.

\sectiono{Summary of the analysis} \label{estrategy}

In this section, we shall outline the general strategy of our analysis.
A first-law type argument reduces the statement of temperature and moduli independence of the index to the temperature independence of the mass and the ``scalar charges" (defined below). 
The asymptotic form of the Killing spinor equations implies this,  since it fixes the mass and the scalar charges in terms of the electric charges,  magnetic charges and asymptotic values of moduli fields,  with no dependence on the temperature.  

Consider a family of classical supergravity solutions with varying $\beta$ and $m^{(a)}$,  but with fixed values of $Q_{(i)}$, $P^{(i)}$ and $\omega$.
From the definition \refb{e4} of $S_{\mathrm{BPS}}$,  we have
\be\label{e5}
\delta S_{\mathrm{BPS}} = -\delta I + M \, \delta\beta  + \beta\, \delta M + Q_{(i)} \, \delta\nu^{(i)}\, 
\ee
as we vary the solution within this family.
%Note that we have kept the asymptotic values of the moduli fields fixed; this is an assumption that we shall be making throughout our analysis.
From \refb{e3}, we have
\be\label{e6}
\delta I = M\, \delta \beta + Q_{(i)}\, \delta\nu^{(i)}  +\beta \, \Sigma_{(a)}\, 
\delta m^{(a)}\, ,
\ee
where we used the fact that $\omega$ is fixed for the entire family.
Substituting this into \refb{e5},  we get
\be\label{e7}
\delta S_{\mathrm{BPS}} =  \beta \,  \delta M -  \beta \, \Sigma_{(a)}\,  \delta m^{(a)}\, .
\ee

We shall show in section \ref{sec-consequences}, using a purely gravitational argument, that,  for a supersymmetric solution,  $M$ and $\Sigma_{(a)}$ are determined in terms of $Q_{(i)}$, $P^{(i)}$ and $m^{(a)}$. 
\emph{In particular,  $M$ and $\Sigma_{(a)}$ are independent of $\beta$},  even though,  
a priori,  the family of solutions under consideration is labelled by both $\beta$ and 
$m^{(a)}$.
For $M$,  this is a restatement of the well-known result that the mass of a supersymmetric black hole is determined in terms of its  charges and the asymptotic moduli (see,  for example,  \cite[section 22.6]{Freedman:2012zz}),\footnote{See also \cite{Ceresole:1995ca,  Hristov:2011qr,  LopesCardoso:2000qm}.  Also note that proving \emph{inequalities} relating mass and charge generally rely on a particular Lagrangian and the behavior of solutions on an entire Cauchy slice,  not just the sphere at infinity \cite{Witten:1981mf,  Nester:1981bjx,  Gibbons:1982fy}.  We are focusing on proving equality between mass and charge for supersymmetric solutions only.} but we shall take care to ensure that the result is not affected by angular momentum or higher-derivative corrections. 
Using this fact and since $Q_{(i)}$ and $P^{(i)}$ are kept fixed for the family of solutions under consideration, we get 
\be
\delta M= {\p M\over \p m^{(a)}}\bigg|_{Q,m} \delta m^{(a)}\, ,
\ee
and hence 
\be\label{e1.11xx}
\delta S_{\mathrm{BPS}}=\beta \left({\p M\over \p m^{(a)}}\bigg|_{Q,m} - \Sigma_{(a)}\right) \, \delta
m^{(a)} \, .
\ee

To proceed further, we define
\be \label{edeftI}
\wt I := I - \nu^{(i)} Q_{(i)}\, .
\ee
Using \refb{e3}  and \refb{edeftI} 
we see that under a general variation of all the parameters,
\be\label{e6xx}
\delta \wt I = M\, \delta \beta - \nu^{(i)} \, \delta Q_{(i)} + J\,  \delta\omega +\beta \, \Sigma_{(a)}\, 
\delta m^{(a)}\, .
\ee
Let us now regard $\wt I$ as a function of $\beta$,  $Q_{(i)}$, $\omega$ and $m^{(a)}$.
It follows that
\be
M = {\p \wt I\over \p\beta}\bigg|_{\beta,Q,\omega,m}, \qquad \beta\Sigma_{(a)} =
{\p \wt I\over \p m^{(a)}} \bigg|_{\beta,Q,\omega,m} \, .
\ee
Using the fact that $M$ and $\Sigma_{(a)}$ are independent of $\beta$
for fixed $Q$, $P$, $m$ and $\omega$,  
we get\footnote{Eq.~\refb{e2.9xx} is close to a similar relation derived in \cite{9607108}, with the only difference that
in \cite{9607108} the derivative with respect to $m^{(a)}$ was taken at fixed area, 
angular momenta
and charges, whereas here the derivative is carried out at fixed $\omega$, $\beta$ and
the charges.}
\be\label{e2.9xx}
{\p M\over \p m^{(a)}}\bigg|_{Q,m} ={\p^2 \wt I\over \p\beta\p m^{(a)}}\bigg|_{\beta,Q,\omega,m}
= {\p\over \p\beta} (\beta \Sigma_{(a)})\bigg|_{\beta,Q,\omega,m}= \Sigma_{(a)}\, .
\ee
Substituting this into \refb{e1.11xx}, we get
\be
\delta S_{\mathrm{BPS}}=0\, .
\ee
This shows that $S_{\mathrm{BPS}}$ is independent of $m^{(a)}$ and $\beta$.

We now describe the set-up in which we shall be working and the underlying assumptions.
\begin{enumerate}
\item We shall work with a general theory of gravity with $\NN\ge 2$ supersymmetry in asymptotically flat four-dimensional spacetime,  without making any assumption about what kind of higher-derivative corrections are present. 
This covers all string compactifications with $\NN\ge 2$ supersymmetry.
\item We shall assume that there is a family of solutions with varying $\beta$ 
and $m^{(a)}$ that preserve at least one supersymmetry, i.e., there is a
spinor such that when we make a supersymmetry transformation with that spinor,
the background remains invariant.
Typically,  these solutions will be single-centered or multi-centered black holes.
\item We assume that, asymptotically, the metric and the gauge fields take the form of a Reissner--Nordstr\"om black hole
up to corrections of order $1/r^2$. Here $r$ is the radial distance.
Note that angular momentum corrects the metric and the gauge fields at order $1/r^2$.
 \item We make no assumption about the structure of the solution in the interior
 except its smoothness.
 \item The solution is in general complex, both in the Euclidean description and in the
 Lorentzian description. However, we shall assume that, in the Lorentzian description, the
 asymptotic form of the solution is real. The effect of having complex angular momentum,
 which arises from having to choose imaginary $\omega$, falls off faster than the order to
 which we need the solution.
\end{enumerate}

We shall now argue that under these assumptions, it is enough to work with the two-derivative $\NN\ge 2$ supergravity part of the action. 
For this, it will be useful to introduce a rescaled coordinate $\sigma$ where $\sigma := \tau/\beta$.
In that case, it follows from \refb{e1}, \refb{e2} that
\be\label{esigma}
(\sigma,\phi) \equiv (\sigma+1, \phi+i\omega), 
\qquad g_{\sigma\sigma} =\beta^2 \quad \hbox{asymptotically}\, .
\ee
Since our goal is to show that $M$ and $\Sigma_{(a)}$ are
determined by $Q_{(i)}$, $P^{(i)}$ and $m^{(a)}$,  we need to argue that these are determined from the asymptotic form of the solution and are insensitive to the higher-derivative corrections.  
The magnetic
charges $P^{(i)}$ are determined from \refb{e1.3} and clearly are defined in terms of the
asymptotic form of the gauge fields. 
Similarly $m^{(a)}$'s are given by the asymptotic values of the scalar fields.
To determine $M$,  we use the definition given in 
\refb{e3}. When we vary $\beta$,  the solution changes and we need to compute the
action of the new solution. Now, since for a solution the first-order variation of the 
Lagrangian density $\LL$ is
a total derivative, the change in the action $I$ will be a boundary term near $r=\infty$. 
This boundary term is best
computed in the $\sigma$ coordinate system introduced in \refb{esigma}.
We compute $\delta\LL / \delta g_{\sigma\sigma}$ and then express it as a 
total derivative
after using the equations of motion. In this computation, the contribution from
any higher-derivative term in $\LL$ 
will be suppressed since, for large $r$, each power of derivative produces a factor of $1/r$.
A similar argument holds for $Q_{(i)}$ and
$\Sigma_{(a)}$ defined through \refb{e3}, with the difference that we now
replace $\delta\LL / \delta g_{\sigma\sigma}$ by $\delta\LL / \delta A^{(i)}_\tau$ and
$\delta\LL / \delta \phi^{(a)}$.

This shows that the relation between $M$, $Q_{(i)}$, $\Sigma_{(a)}$ 
and the asymptotic form of the fields does
not get affected by higher-derivative corrections to the action. 
We shall further show in section \ref{sec-consequences} that the existence of a Killing spinor relates the asymptotic
form of the fields, and this determines $M$ and $\Sigma_{(a)}$
in terms of $Q_{(i)}$, $P^{(i)}$ and $m^{(a)}$. 
The supersymmetry transformation laws, and hence the  Killing
spinor equations, also get modified by higher-derivative corrections, but these 
corrections are also
suppressed in the asymptotic region.
Hence,  the
relation between $M$, $\Sigma_{(a)}$ and 
$Q_{(i)}$, $P^{(i)}$ and $m^{(a)}$ is not affected by higher-derivative corrections, and
we can
restrict our analysis to two-derivative supergravity.\footnote{The upcoming paper \cite{luca-to-appear} computes the on-shell action in $\mathcal{N}=2$ SUGRA with an arbitrary number of higher derivatives and finds by explicit computation that the mass is independent of $\beta$. A related analysis
for AdS black holes can be found in \cite{1810.11442}.}

Finally, note that the argument holds for each classical saddle solution separately.
For example, $\NN=2$ supergravity is known to have multi-centered 
black hole solutions
that contribute to the index \cite{2507.07166}. 
We can define
$S'_{\mathrm{BPS}}$ to be the quantity $S_{\rm BPS}$ defined in  \refb{e4} for this
solution.\footnote{Note that, at the end, 
$e^{S_{\mathrm{BPS}}}$ for each saddle, multiplied
by  the result of the path integral along the Lefschetz thimble of the saddle, needs to be added
to get the total index.} Then the same arguments as given above will tell us that $S'_{\mathrm{BPS}}$
is independent of $\beta$ and $m^{(a)}$.

\sectiono{Definitions of mass and charge}

In the asymptotic region we can express the metric and the gauge fields as 
the Reissner--Nordstr\"om metric in isotropic coordinates $(t,x^1,x^2,x^3)$ as follows
\ben \label{esolution}
ds^2 &=& -g(r)^2 dt^2 + f(r)^2 \, \delta_{mn} dx^m dx^n, \qquad r = \sqrt{\delta_{mn} x^m x^n},
\qquad 1\le m,n\le 3\, ,
\nonumber \\
&& 
F^{(i)}_{rt} = {E^{(i)}\over 4\pi\, r^2}, \qquad F^{(i)}_{\theta\phi} = {P^{(i)}\over 4\pi} \sin\theta, \qquad
\phi^{(a)} = m^{(a)} - {\sigma^{(a)}\over r}\, ,
\een
\be \label{egrexp}
g(r) = 1- K/r+\OO(r^{-2}), \qquad f(r) = 1 + K/r + \OO(r^{-2})\, .
\ee
Here $K$ is related to the mass of the solution,  see (\ref{ekgm}).  Note that the coefficient $E^{(i)}$ appearing in the asymptotic form of $F^{(i)}_{rt}$ is not necessarily the same as the electric charge $Q^{(i)}$.
Note that even though the solution carries angular momentum, its effect on the
asymptotic metric is subleading, contributing a term of order $1/r^3$ to the spin
connection. We shall need the spin connection up to terms of order $1/r^2$ and hence 
we can ignore that effect of the angular momentum.

We will denote the tangent space indices by $a,b$ and choose the vierbeins as
\be \label{e2.3}
e_t^0=g(r), \qquad e_m^a = f(r)\, \delta_m^a, \qquad 1\le m\le 3, \quad 0\le a,b\le 3\, .
\ee
The inverse vierbeins are
\be
E_0^t = g(r)^{-1}, \qquad E_1^x = E_2^y=E_3^z = f(r)^{-1}\, .
\ee

If we define the dual field strength
\be
\wt F^{(i)\mu\nu} ={1\over 2} \, \eps^{\mu\nu\rho\sigma} \, F^{(i)}_{\rho\sigma}\, ,
\ee
then we get
\be
\wt F^{(i)}_{rt} = {P^{(i)}\over 4\pi r^2}\, .
\ee
Our convention for the $\epsilon$ tensor is $\eps^{tr\theta\phi} =1/\sqrt{-\det g}$.

After making the Wick rotation $t\to -i\tau$, the metric takes the form
\be
ds^2 = g(r)^2 d\tau^2 + f(r)^2 (dx^2 + dy^2 + dz^2)\, .
\ee
The Euclidean action $I$ is given by
\be
I = -\int d\tau\, \int d^3 x\, \sqrt{\det g} \, \LL\, .
\ee
We can calculate the mass and the charges using the definitions given in \refb{e3}.
For calculating the mass we add the Gibbons--Hawking boundary term and vary the
action with respect to $g_{\sigma\sigma}$ defined in \refb{esigma}.
 This gives us back the standard
result that $K$ appearing in \refb{egrexp} is related to the mass by the relation
\be\label{ekgm}
K = GM\, ,
\ee
where $G$ is Newton's constant. This relation is not affected by the addition of
higher-derivative terms in the action since their contribution falls off faster as $r\to\infty$ and
hence they do not contribute to the boundary terms.

To compute the charges $Q_{(i)}$, we shall vary the asymptotic value of 
the gauge fields $A^{(i)}_\tau$ 
and pick up the boundary term at $r=\infty$. The relevant term takes the form
\ben\label{echange}
&& - 2\,  \int d\tau\, \int d^3 x\, \sqrt{\det g} \  
{\delta \LL\over \delta F^{(i)}_{r\tau}} \delta F^{(i)}_{r\tau}
\nonumber \\
&=&-2\, 
 \int d\tau\, \int d^3 x\,  \p_r \left[\sqrt{\det g} \ 
{\delta \LL\over \delta F^{(i)}_{r\tau}} \delta A^{(i)}_{\tau} \right] + \hbox{terms
that vanish by the equations of motion} \nonumber \\
&=& -8 \pi r^2 \, \beta\,  {\delta \LL\over \delta F^{(i)}_{r\tau}}  
(-i\, \delta \nu^{(i)}/\beta )\Bigg|_{r=\infty}\, .
\een
The first factor of 2 comes from having to sum over variations of $F^{(i)}_{r\tau}$ and
$F^{(i)}_{\tau r}$.
Given the form of $\LL$ we can compute $\delta \LL/\delta F^{(i)}_{r\tau}$ and
express this in terms of $E^{(i)}$ and $P^{(i)}$ appearing in \refb{esolution}. 
Some important
points to note are as follows:
\begin{enumerate}
\item The $\beta$ dependence from the $\tau$ integration range cancels the
$1/\beta$  from the expression for $\delta A^{(i)}_\tau$. As a result the expression
depends only on the asymptotic values of field strengths parametrized by $E^{(i)}$ and
$P^{(i)}$ and the asymptotic values of various scalar fields but not on $\beta$.
\item The $r^2$ from the surface area cancels the $1/r^2$ factors coming from the field
strength.
\end{enumerate}
Identifying $Q_{(i)}$ as the coefficient of the $\delta\nu^{(i)}$ term in \refb{echange},
\be
Q_{(i)}= 8\, \pi\, i\,  r^2 \,  {\delta \LL\over \delta F^{(i)}_{r\tau}}  
\Bigg|_{r=\infty}\, ,
\ee 
we can
express $Q_{(i)}$ as a function of $E^{(j)}$, $P^{(j)}$ and the asymptotic values of the scalar
fields. The relation is not affected by the addition of higher-derivative terms in the action
since their contribution falls off faster than $1/r^2$. We can invert this to find
$E^{(i)}$ in terms of $Q_{(j)}$, $P^{(j)}$ and $m^{(a)}$.

Finally, to compute the scalar charges $\Sigma_{(a)}$, we shall vary the 
asymptotic value of 
the scalar fields $\phi^{(a)}$ 
and pick up the boundary term at $r=\infty$. The relevant term is
\ben
&& -\int d\tau\, \int d^3 x\, \sqrt{\det g} \  
{\delta \LL\over \delta \p_r\phi^{(a)}} \delta \p_r \phi^{(a)}
\nonumber \\
&=& 
- \int d\tau\, \int d^3 x\,   \p_r \left[\sqrt{\det g} \
{\delta \LL\over \delta \p_r\phi^{(a)}} \delta \phi^{(a)} \right] + \hbox{terms
that vanish by the equations of motion} \nonumber \\
&=& -4\pi r^2 \, \beta\,  {\delta \LL\over \delta (\p_r\phi^{(a)})}  
\delta m^{(a)}\Bigg|_{r=\infty}\, .
\een
Comparing this with \refb{e3} we get
\be \label{e2.13xx}
\Sigma_{(a)} = -4\pi r^2 \, {\delta \LL\over \delta (\p_r\phi^{(a)})}  
\Bigg|_{r=\infty}\, .
\ee
This determines $\Sigma_{(a)}$ in terms of $\sigma^{(a)}$ and $m^{(a)}$.

One point to be noted is that as long as the Lagrangian density is real and the
real moduli fields approach real values at infinity, $E^{(i)}$ is a real function of 
$Q_{(j)}$, $P^{(j)}$ and $m^{(b)}$. Similarly,  
$\Sigma_{(a)}$ is a real function of $\sigma^{(b)}$ and
$m^{(b)}$.

\sectiono{Supersymmetry transformations}

The solutions that we shall need to analyze are in general
complex both in Euclidean and Lorentzian
signature. We shall work in the Lorentzian signature and first write down the
Killing spinor equations  in terms of real fields and real transformation 
parameters. Then we shall allow complex solutions by allowing the fields to be
complex and also the supersymmetry transformation parameters to be complex.

Usually the supersymmetry transformations are written in terms of complex fields.
They take the form~(see, e.g., \cite[sections 3.2 and 4]{Andrianopoli:1996ve} and \cite[section 3.2]{Andrianopoli:2006ub})
\be \label{e3.1}
\delta \psi_{\mu A} = D_\mu \eps_A +  i\, T^-_{\mu\nu AB} \, E_a^\nu \, \gamma^a \,
\eps^B
= \p_\mu \eps_A + {1\over 4} \, \omega_\mu^{ab} \, \gamma^{ab} 
\eps_A + i\, T^-_{\mu\nu AB}  \, E_a^\nu \, \gamma^a \, \eps^B\, .
\ee
Here $\omega_\mu^{ab}$ is the spin connection, $T_{\mu\nu AB}$ 
are the graviphoton field strengths given by appropriate complex linear combinations of
the various real electromagnetic fields $F^{(i)}_{\mu\nu}$ of the theory and their duals
$\wt F^{(i)}_{\mu\nu}$
with coefficients given by the scalar moduli fields of the theory.
$T^-_{\mu\nu AB}$ is the anti-self-dual part of $T_{\mu\nu AB}$:
\be
T^-_{\mu\nu AB} = {1\over 2} (T_{\mu\nu AB} - i\, \wt T_{\mu\nu AB} )\, .
\ee
The indices $A,B$ run over $\NN$ values in a theory with $\NN$ supersymmetries.
$T_{\mu\nu AB}$ is anti-symmetric under the exchange of $A$ and $B$.
$\eps_A$ and $\psi_{A\mu}$ are chiral spinors, satisfying
\be \label{e3.4}
\gamma^5 \eps_A = \eps_A, \qquad \gamma^5 \psi_{A\mu} = \psi_{A\mu}\, ,
\ee
where
\be
\gamma^5 := i\, \gamma^0\gamma^1\gamma^2\gamma^3\, .
\ee
$\eps^A$ are charge conjugates of $\eps_A$ of opposite chirality. In 
the Majorana representation where all $\gamma$ matrices are purely real,
we have
\be 
\eps^A = (\eps_A)^*, \qquad \gamma^5\eps^A = - \eps^A\, .
\ee
We shall use the Majorana representation for our analysis.

Setting the right-hand side of \refb{e3.1} to zero, we get the
Killing spinor equation
\be \label{em1}
D_\mu \eps_A +  i\, T^-_{\mu\nu AB} E_a^\nu \gamma^a \eps^B = 0\, .
\ee

We shall work with real fields by expressing every complex field in terms of its real
and imaginary parts. We write
\be \label{em2}
T_{\mu\nu AB}=T_{R;\mu\nu AB} + i\, T_{I;\mu\nu AB}, \qquad
T^-_{\mu\nu AB} = {1\over 2}\left(T_{R;\mu\nu AB} + \wt T_{I;\mu\nu AB} - i \wt T_{R;\mu\nu AB} 
+ i\, T_{I;\mu\nu AB}\right)\, ,
\ee
and
\be \label{em3}
\psi_{\mu A} = \psi_{R;\mu A} + i\, \psi_{I; \mu A}, \qquad \eps_A = \eps_{R; A} + i\, \eps_{I; A},
\qquad \eps^A = \eps_{R; A} - i\, \eps_{I; A}\, .
\ee
Since $\gamma^5$ is purely imaginary, we get, from \refb{e3.4},
\be
\gamma^5 \, \eps_{R;A} = i \eps_{I;A}, \qquad \gamma^5 \, \eps_{I;A} =- i \eps_{R;A}\, .
\ee

Eq.\refb{em1} now gives
\ben \label{e3.1R}
\p_\mu \eps_{R;A} + {1\over 4} \, \omega_\mu^{ab} \, \gamma^{ab} 
\eps_{R;A} +  G_{\mu\nu AB}  \, E_a^\nu \, \gamma^a \,\eps_{R;B} +  H_{\mu\nu AB}  \, E_a^\nu \, \gamma^a \,\eps_{I;B} =0\, ,
\een
and
\ben \label{e3.1I}
\p_\mu \eps_{I;A} + {1\over 4} \, \omega_\mu^{ab} \, \gamma^{ab} 
\eps_{I;A}  - G_{\mu\nu AB} \, E_a^\nu \, \gamma^a \, \eps_{I;B} +
H_{\mu\nu AB}   \, E_a^\nu \, \gamma^a \,\eps_{R;B}=0 \, ,
\een
where
\be\label{edefGH}
H_{\mu\nu AB} = {1\over 2} \left\{ T_{R; \mu\nu AB} +  \wt T_{I;\mu\nu AB}
\right\} , \qquad G_{\mu\nu AB} = {1\over 2}  \left\{ 
\wt T_{R; \mu\nu AB} - T_{I; \mu\nu AB} 
\right\}  \, .
\ee
Since the $\gamma^a$'s are real matrices,
all terms in \refb{e3.1R}, \refb{e3.1I} are real for real fields.

For a complex solution, the
fields $T_{R;\mu\nu AB}$, $T_{I;\mu\nu AB}$ and the metric are complex.
In this case we need to allow the 
transformation parameters $\eps_{R; A}$ and $\eps_{I; A}$ to be complex to find the
Killing spinors. Since $T_{R;\mu\nu AB}$ and $T_{I;\mu\nu AB}$ are determined in terms
of the electric and magnetic fields and the moduli, $G_{\mu\nu AB}$ and $H_{\mu\nu AB}$
are also determined in terms of the same fields. Hence the asymptotic values of
$G_{\mu\nu AB}$ and $H_{\mu\nu AB}$ are determined solely in terms of the electric
and magnetic charges $P^{(i)}$ and $Q_{(i)}$ and the asymptotic values of the moduli.
These, in turn, remain real since $P^{(i)}$ and $Q_{(i)}$ are real.

Since $\omega_\mu^{ab}$ and the field strengths $G_{\mu\nu AB} $ and
$H_{\mu\nu AB}$ fall off as $1/r^2$ for large $r$, we can immediately conclude from
\refb{e3.1R}, \refb{e3.1I} that 
asymptotically $\eps_{R;A}$ and $\eps_{I;A}$ must be constant, independent
of the angular and radial coordinates. 
Furthermore for a stationary solution of the type we are considering, $\eps$ is
time independent.
We shall use this information in the
analysis of the next section.

We also need the transformation laws of the matter fermions. In the classification based on
$\NN=2$ supersymmetry, these include
the gauginos of the vector multiplets and the
hyperinos. They transform as~(see, e.g., \cite[sections 3.2 and 4]{Andrianopoli:1996ve} and \cite[section 3.2]{Andrianopoli:2006ub}, and, for the hyperinos, \cite[section 4]{Andrianopoli:1996vr})
\be \label{ematterfermion}
\delta\chi^{P} = \SSS^{P}{}_{(a)A}\, (\p_\mu \phi^{(a)})\, E^\mu_b\, \gamma^b\, \eps^A
+ \MM^{P\, A -}_{\mu\nu}\, E^\mu_b\, E^\nu_c\, \gamma^{bc}\, \eps_A
+ (\hbox{terms involving fermions})\, ,
\ee
where $P$ is a collective index running over the matter fermions, $\SSS^P{}_{(a)}{}_A$ is
a moduli-dependent coefficient, and $\MM^{P\,A-}_{\mu\nu}$ is the anti-self-dual part
of the combinations of the field strengths orthogonal to the graviphoton, with
moduli-dependent coefficients. The precise index structure depends on $\NN$, but the
two features we need are common to all theories: the variation is linear in
$\p_\mu \phi^{(a)}$ and linear in the matter field strengths. 
Setting the real and imaginary parts of \refb{ematterfermion} to
zero we get
\ben \label{emodfin}
&& M^{P}{}_{(a)}{}^{A}\, (\p_\mu \phi^{(a)})\, E^\mu_b\, \gamma^b\, \eps_{R;A}
+ N^{P}{}_{(a)}{}^{A}\, (\p_\mu \phi^{(a)})\, E^\mu_b\, \gamma^b\, \eps_{I;A}
+ R^{P\, A}_{\mu\nu}\, E^\mu_b\, E^\nu_c\, \gamma^{bc}\, \eps_{R;A}
\nonumber \\ && \hskip 1in 
+ S^{P\, A}_{\mu\nu}\, E^\mu_b\, E^\nu_c\, \gamma^{bc}\, \eps_{I;A}=0,
\nonumber \\
&& U^{P}{}_{(a)}{}^{A}\, (\p_\mu \phi^{(a)})\, E^\mu_b\, \gamma^b\, \eps_{R;A}
+ V^{P}{}_{(a)}{}^{A}\, (\p_\mu \phi^{(a)})\, E^\mu_b\, \gamma^b\, \eps_{I;A}
+ W^{P\, A}_{\mu\nu}\, E^\mu_b\, E^\nu_c\, \gamma^{bc}\, \eps_{R;A}
\nonumber \\ && \hskip 1in 
+ X^{P\, A}_{\mu\nu}\, E^\mu_b\, E^\nu_c\, \gamma^{bc}\, \eps_{I;A}=0\, ,
\een
where $M^{P}{}_{(a)}{}^{A},\cdots , X^{P\, A}_{\mu\nu}$ are real functions of real
moduli fields and gauge field strengths.

\sectiono{Consequences of the Killing spinor equation}
\label{sec-consequences}

In this section,  we complete the argument of section \ref{estrategy} by showing that the asymptotic form of the Killing spinor equations determines the mass $M$ and the scalar charges $\Sigma_{(a)}$ in terms of $Q_{(i)},  P^{(i)}$ and $m^{(a)}$.
In particular,  $M$ and $\Sigma_{(a)}$ are independent of $\beta$.

It follows from the $\mu=t$ component of the Killing spinor equations 
\refb{e3.1R}, \refb{e3.1I} that we have
\ben \label{e4.2}
&& \lim_{r\to\infty} r\, \, \vec x.\vec\gamma \, \left({1\over 4} \, \omega_t^{de} \, \gamma^{de} 
\eps_{R;A}  +   G_{0 m AB}  \, \gamma^m \,\eps_{R;B} + H_{0m AB}  \,  \gamma^m 
\,\eps_{I;B} 
\right)
=0, \nonumber \\ 
&& \lim_{r\to\infty} r \, \vec x.\vec\gamma \,   \left(
 {1\over 4} \, \omega_t^{de} \, \gamma^{de} 
\eps_{I;A}  -  G_{0m AB}  \,  \gamma^m \,\eps_{I;B} + H_{0m AB}  \, \gamma^m 
\,\eps_{R;B} 
\right)
=0\, .
\een
Note that we have multiplied the Killing spinor equation by $\vec x.\vec\gamma$.
For the vierbeins given in \refb{e2.3}, we have
\be
\omega_t^{0a} = {g'\over f} \, {x^a\over r}\, , \qquad 1\le a,b,c\le 3\, .
\ee
This gives, using \refb{egrexp}, \refb{ekgm},
\be \label{exx1}
{r\over 4} \,  \vec x.\vec\gamma \,  \omega_t^{de} \, \gamma^{de} 
\eps = {1\over 2} {g'\over f} \, \vec x.\vec\gamma \, \gamma^0 \,  \vec x. \vec\gamma \, \eps
=-{1\over 2} \, G\, M\, \gamma^0  \, \eps\, .
\ee

For evaluating the terms involving $G_{\mu\nu AB}$ and $H_{\mu\nu AB}$
in \refb{e4.2} we first note that the asymptotic forms of these fields are given by
\be \label{exx2}
G_{0m AB} = \GG_{AB} \, {x^m\over r^3}, \qquad H_{0m AB} = \HH_{AB} \, {x^m\over r^3}\, ,
\ee
where $\GG_{AB}$ and $\HH_{AB}$ are determined in terms of the charges 
$Q_{(i)}$, $P^{(i)}$ and the moduli.
Using \refb{exx1}, \refb{exx2} we can express \refb{e4.2} as
\ben
-{1\over 2} \, G\, M\, \gamma^0  \, \eps_{R;A}  + \GG_{AB} \eps_{R;B} + \HH_{AB}
\eps_{I;B}=0, \nonumber \\
-{1\over 2} \, G\, M\, \gamma^0  \, \eps_{I;A}  - \GG_{AB} \eps_{I;B} +
\HH_{AB}
\eps_{R;B}=0\, .
\een
This can be written as
\be\label{e5.5zyx}
-GM\begin{pmatrix}\gamma^0 & 0 \\ 0\, & \gamma^0\end{pmatrix} \begin{pmatrix}\eps_{R;A}\\ \eps_{I;A}\end{pmatrix}
+  2 \begin{pmatrix} \GG_{AB} &  \HH_{AB} \\
\HH_{AB} &  
-\GG_{AB}\end{pmatrix} \begin{pmatrix}\eps_{R;B}\\ \eps_{I;B}\end{pmatrix}=0\, .
\ee
This has a solution only if the associated matrix
has vanishing determinant:
\be \label{edeteq}
\det \begin{pmatrix} - G M \gamma^0 +2\, \GG & 2\,  \HH \\
2\,  \HH & -GM\gamma^0-2\, 
\GG\end{pmatrix} =0\, .
\ee
This determines $M$ as a function of the charges and moduli without any dependence on
$\beta$.\footnote{Eq.~\refb{edeteq} has multiple roots, but along any given branch the solution is
independent of $\beta$.
}

\def\WWW{{\cal W}}

We can analyze the consequence of \refb{emodfin} in a similar manner.
Asymptotically, from \refb{esolution},
\be
\p_m\, \phi^{(a)} =   \sigma^{(a)}\, {x^m\over r^3}\, , 
\ee
and the $0m$ and $mn$ components of $R,S,W,X$ all scale as $1/r^2$
with coefficients given by functions of $Q_{(i)}$, $P^{(i)}$ and $m^{(a)}$. 
Hence by setting
the coefficient of $1/r^2$ to zero we get an equation of the form
\be \label{eWP}
\WWW^P \begin{pmatrix}\eps_{R}\\ \eps_{I}\end{pmatrix}=0\, ,
\ee
where $\eps_R$ and $\eps_I$ are column matrices labelled by $A$ and the Dirac
index and $\WWW^P$ are matrices that depend linearly on $\sigma^{(a)}$ and
the charges $P^{(i)}$ and $Q_{(i)}$ and non-linearly on the moduli $m^{(a)}$.
The period $\beta$ enters
nowhere. 
Requiring that these equations, together with \refb{e5.5zyx}, have a common
non-zero solution
gives a
set of algebraic relations that determine
$\sigma^{(a)}$ in terms of $Q_{(i)}$, $P^{(i)}$ and $m^{(a)}$:\footnote{This 
argument assumes that all $\sigma^{(a)}$'s are determined from these equations. 
This could fail if some linear combination $c_{(a)}\p_\mu\phi^{(a)}$ did not enter 
\refb{ematterfermion}. 
This would imply that we could get a supersymmetric
configuration if we switched on a
non-zero $\sum_{a} c_{(a)}\vec\nabla \phi^{(a)}$. But this should not
be possible since such a configuration will have non-zero energy density.}
\be \label{eMsigma}
 \sigma^{(a)} = \sigma^{(a)}(Q_{(i)},P^{(i)},m^{(b)})\, ,
\ee
with no dependence on $\beta$. 
\refb{e2.13xx} now determines $\Sigma_{(a)}$  
in terms of $Q_{(i)}$, $P^{(i)}$ and $m^{(a)}$.

By the
argument of section \ref{estrategy}, the corrections to \refb{ematterfermion} from higher-derivative
terms in the action carry additional powers of derivatives and hence fall off faster
than $1/r^2$, so \refb{eWP} is uncorrected.

This establishes that $M$ and $\Sigma_{(a)}$ are functions of $Q_{(i)}$, $P^{(i)}$ 
and $m^{(b)}$ but are independent of $\beta$.
The argument given in section \ref{estrategy} then
shows that the logarithm of the 
index, $S_{\rm BPS}$, defined in \refb{e4}, is independent of $\beta$ and
$\{m^{(a)}\}$.
For the specific question of moduli independence of the index,  our result provides a purely asymptotic argument,  in contrast to the attractor mechanism \cite{9508072,9602111,9602136,9603090}.

We have only proved the temperature and moduli independence of the index in the classical limit on the gravity side.  We leave it as an interesting future direction to see if the argument can be extended to one-loop order and beyond using a purely gravitational analysis.
It would also be interesting to work out the details of the proof in the asymptotically AdS case both classically and at higher-loop orders.

\paragraph{Acknowledgments.}
This work was supported by the Department of Atomic Energy, Government of India, under project no.~RTI4019.
We would like to thank J. Boruch,  Y.  Chen,  L.  Iliesiu, S. Murthy, D. Stanford and
G.  J.  Turiaci for discussions.
RM thanks the Leinweber Institute for Theoretical Physics at Stanford for its hospitality during the completion of part of this work.
The work of A.S. was supported by the ICTS-Infosys Madhava Chair Professorship. 

\paragraph{Statement on AI use.} The arguments presented in this paper were developed with assistance from AI tools (ChatGPT,  Claude and an internal model at Google DeepMind).
The manuscript was written by the authors,  who take full responsibility for any errors.

\bibliographystyle{apsrev4-1long}
\bibliography{beta}

\end{document}